\documentclass[
 reprint,
 superscriptaddress,
 amsmath,amssymb,
 aps,prl
 ]{revtex4-2}

\usepackage{graphicx}
\usepackage{dcolumn}
\usepackage{bm}

\usepackage{subfloat}
\usepackage[usenames]{color}

\newcommand{\be}{\begin{equation}}
\newcommand{\ee}{\end{equation}}
\newcommand{\ba}{\begin{eqnarray}}
\newcommand{\ea}{\end{eqnarray}}
\newcommand{\bd}{\begin{displaymath}}
\newcommand{\ed}{\end{displaymath}}

\def\thalf{{\textstyle{\frac{1}{2}}}}

\def\oneqt{{\textstyle{\frac{1}{4}}}}

\begin{document}

\preprint{APS/123-QED}

\title{Disoriented isospin condensates may be the source of anomalous kaon correlations measured in Pb-Pb collisions at $\sqrt{s_{NN}} = 2.76$ TeV}
 
\author{Joseph I. Kapusta}
 \affiliation{School of Physics \& Astronomy, University of Minnesota, Minneapolis, MN 55455, USA}

\author{Scott Pratt}
 \affiliation{Department of Physics and Astronomy and Facility for Rare Isotope Beams,\\
Michigan State University, East Lansing, MI 48824, USA}

\author{Mayank Singh}
 \affiliation{School of Physics \& Astronomy, University of Minnesota, Minneapolis, MN 55455, USA}
 

\begin{abstract}
The magnitude of fluctuations between charged and neutral kaons measured by the ALICE Collaboration in heavy-ion collisions at the LHC exceeds conventional explanation. Here it is shown that if the scalar condensate, which is typically associated with chiral symmetry, is accompanied by an isospin=1 field, then the combination can produce large fluctuations where $\langle \bar{u}u\rangle \ne \langle \bar{d}d\rangle$. Hadronizing strange and anti-strange quarks might then strongly fluctuate between charged ($u\bar{s}$ or $s\bar{u}$) and neutral ($d\bar{s}$ or $s\bar{d}$) kaons.
\end{abstract}


\maketitle

The search for experimental manifestation of coherent fields related to the QCD chiral phase transition has attracted great attention in the field of relativistic heavy ion physics. Despite the fact that lattice gauge theory shows that the scalar quark-antiquark condensates should decrease and approach zero for temperatures above 150 MeV, no strong experimental evidence has yet been found. Extensive searches for novel fluctuations of the scalar quark-antiquark field due to rotation with pion fields, known as the disoriented chiral condensate (DCC) \cite{Anselm1,Blaizot:1992at,Rajagopal:1992qz,Bjorken:1993cz,Anselm2,reviewA}, also found no evidence \cite{WA98:1997gpd,NA49:1999inh,MiniMax:1999mxv,WA98:2000zdn,WA98:2002pde}. For a history see Refs. \cite{QMseries,Harris_Muller}.
Whereas the DCC search involved measurements of fluctuations of neutral vs. charged pions, the ALICE Collaboration has recently published results for fluctuations of charged vs. neutral kaons for Pb+Pb collisions at $\sqrt{s_{NN}} = 2.76$ TeV \cite{ALICE:2021fpb}. The magnitude of the measurement well exceeded expectations from typical effects such as charge conservation or Bose-Einstein symmetrization \cite{Kapusta:2022ovq}. The observable analyzed by ALICE is known as $\nu$-dynamic,
\begin{eqnarray}
\nu_{\rm dyn}&=&
  \frac{\langle N_\pm(N_\pm-1)\rangle}{\langle N_\pm\rangle^2}
+ \frac{\langle N_S(N_S-1)\rangle}{\langle N_S\rangle^2}\\
\nonumber&&\hspace*{10pt}
-2\frac{\langle N_SN_\pm\rangle}{\langle N_S\rangle\langle N_\pm\rangle}.
\end{eqnarray}
Here $N_\pm$ refers to the number of observed charged kaons and $N_S$ refers to the number of observed neutral kaons, which in this case are only $K_S$ mesons because ALICE does not measure $K_L$ mesons. To scale the result in such a way that cancels the usual inverse scaling of correlations with multiplicity, ALICE presents $\nu_{\rm dyn}$ scaled by the factor $\alpha=(\langle N_\pm\rangle+\langle N_S\rangle)/\langle N_\pm\rangle\langle N_S\rangle$. For the most central collisions, ALICE's measurement of $\nu_{\rm dyn}/\alpha \approx 0.2$ is about five times the expected value, which mostly derives from charge conservation effects.  This suggests that in the most central collisions, for every charged kaon, one expects approximately 0.25 extra (beyond usual expectations) charged kaons to be distributed in the acceptance region, and for every neutral kaon observed, one expects approximately 0.25 additional neutral kaons to be emitted in the acceptance region. Further, the additional charged, or neutral, kaon production appears to be spread uniformly in rapidity. Thus, if some strange/anti-strange quark binds with an up/anti-up quark to form a charged kaon, it seems that another strange quark will be inspired to also bind with an up/anti-up quark, rather than with a down/anti-down quark to form a neutral meson, even if the two kaons are well separated in rapidity. 

The correlation observed by ALICE is anomalous in its magnitude, in its extent in rapidity, and in its multiplicity dependence. In reference \cite{Kapusta:2022ovq} the usual causes of correlation were studied, namely those from local charge conservation, particle decays, or from Bose-Einstein effects. The measured value of $\nu_{dyn}$ was found to be close to expected values for peripheral collisions, but the measurement well exceeded such expectations for the most central collisions. For any local correlation, the amount of correlation contributed by a given kaon should saturate with increasing volume, or increasing centrality. The observation that $\nu_{dyn}/\alpha$ continually increases with the overall multiplicity, and that that correlation extends across a large rapidity range, must be driven by correlations that extend over a correspondingly large domain size. This strongly suggests a mechanism where correlations are seeded at early times, so that the expanding system can share the correlation across a large spatial rapidity interval. Naturally this leads one to consider mean fields or condensates, given that they might be rapidly undergoing profound transformations, and might also be characterized by large fluctuations. If the seeds of such fluctuations are set at early times, the fluctuations could conceivably result in large domains spanning a wide swath in rapidity. Such correlations might then also grow with increasing centrality.

The purpose of this letter is to investigate whether this correlation could be explained by event-by-event fluctuations of the $\langle \bar{u}u \rangle$ vs. the 
$\langle \bar{d}d \rangle$ condensates. We refer to this as a disoriented isospin condensate (DIC). One could imagine scalar condensates of any of the four forms
$\langle \bar{u}u \rangle,  \langle \bar{d}d \rangle, \langle \bar{u}d \rangle, \langle \bar{d}u \rangle$.
Three of these form an $I=1$ iso-triplet: $\langle \bar{d}u \rangle, (\langle \bar{u}u \rangle - \langle \bar{d}d \rangle)/\sqrt{2}$, and $\langle \bar{u}d \rangle$.  The lowest excitations are associated with the $a_0^+$, $a_0^0$, and $a_0^-$ mesons, the $a_0(980)$.  
The fourth combination $(\langle \bar{u}u \rangle + \langle \bar{d}d \rangle)/\sqrt{2}$
forms an iso-singlet.  Its lowest excitation is associated with the $f_0(500)$, or $\sigma$ meson.  If only the iso-singlet field were present, it should couple equally to charged and neutral kaons. However, if both the $I=1,I_3=0$ and $I=0,I_3=0$ contributions are present in similar amounts, the combination could combine to form nearly all $\langle \bar{u}u \rangle$ or all $\langle \bar{d}d \rangle$ condensates. Assuming that the relative signs were random, this would result in correlated creation of mesons containing of up/anti-up quarks vs. those containing down/anti-down quarks, with the correlation extending over the domain where the fields had correlated signs. One can consider the four-dimensional space comprised of the iso-singlet and iso-triplet scalar fields. The vacuum condensate should be the iso-singlet, but if the free-energy barriers are small, one could easily fluctuate into the iso-triplet directions, thus providing fluctuations in 
$\langle \bar{u}u \rangle$ vs. $\langle \bar{d}d \rangle$.
Assuming that the dissolution of the $\langle \bar{u}u \rangle$ condensate results in charged kaons, and the that dissolution of the $\langle \bar{d}d \rangle$ condensate produces neutral kaons, the fluctuation of these scalar condensates would directly translate into the kaon flavor fluctuations described by $\nu_{\rm dyn}$. The DIC mechanism differs from the coherent mechanisms discussed in \cite{Gavin:2001uk,Kapusta:2022ovq,Nayak:2019qzd}, which involved coherent emission from fields with quantum numbers related to kaon fields, which are rather difficult to motivate.

\begin{figure}
\centerline{\includegraphics[width=0.48\textwidth]{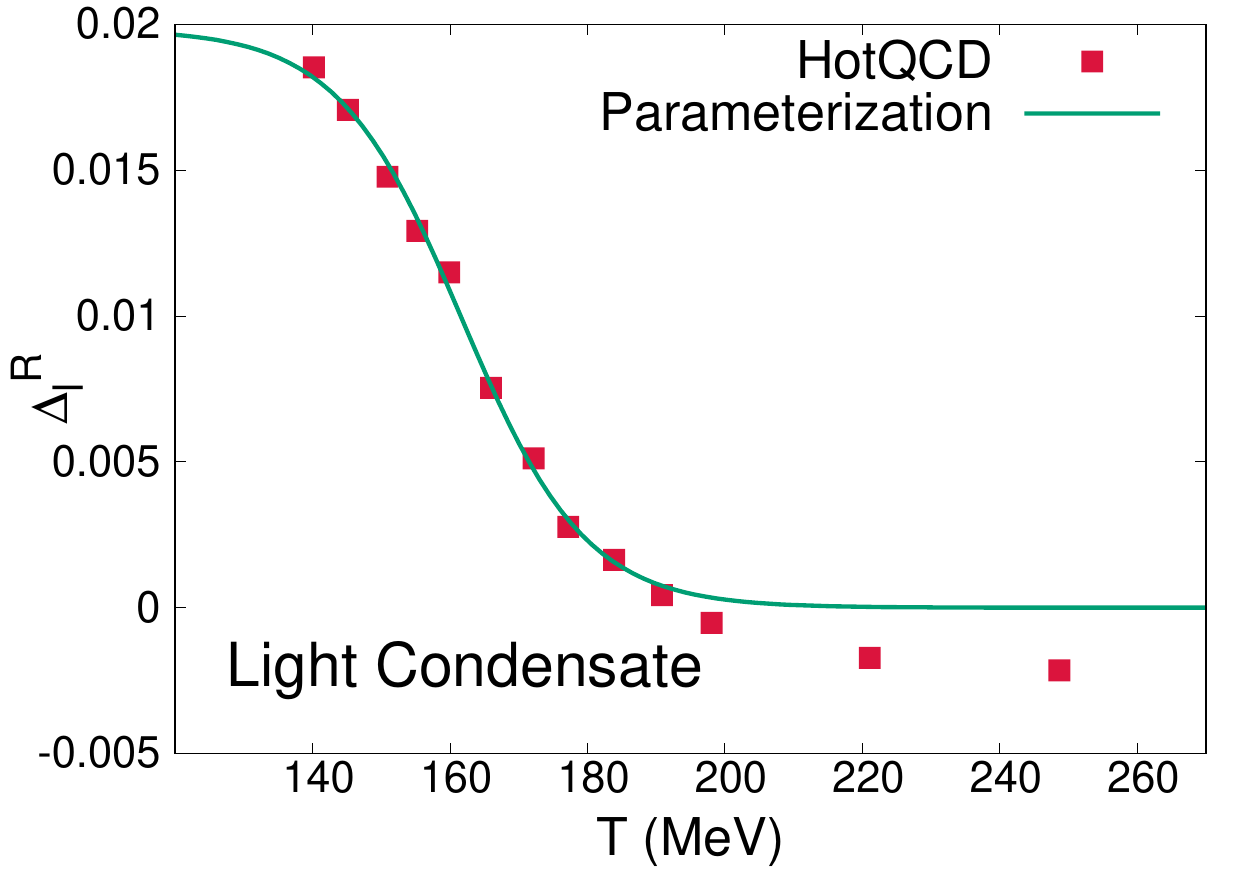}}
\centerline{\includegraphics[width=0.48\textwidth]{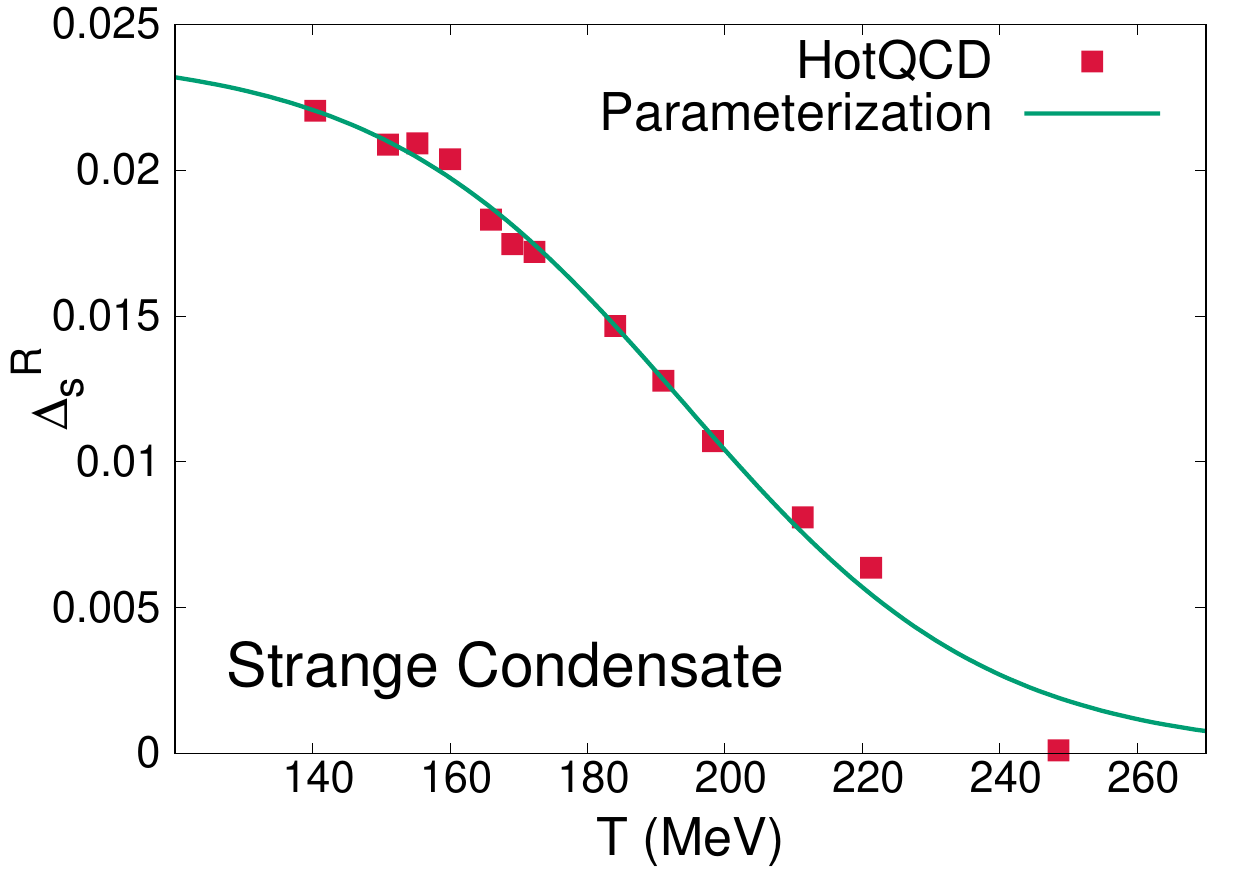}}
\caption{
The quark condensates in dimensionless lattice units vs. temperature as calculated by the Hot QCD Collaboration \cite{HotQCD1}. The condensates are described by the form 
$A/[e^{(T-T_0)/\Delta T}+1]$. The parameters are 
$A=0.01984$, $T_0=161.7$ MeV and $\Delta T=9.009$ MeV 
for the light quark condensate and 
$A=0.02402$, $T_0=194.0$ MeV and $\Delta T=22.25$ MeV
for the strange quark condensate.
}
\label{fig:udsvsT}
\end{figure}
It is easy to motivate the existence of the iso-singlet scalar field, as it is related to chiral symmetry breaking, which tends to be restored for temperatures exceeding the hadronization temperature $T_h\approx 155$ MeV.  Figure \ref{fig:udsvsT} shows the temperature dependence of the quark condensates as calculated by the HotQCD Collaboration \cite{HotQCD1}.  The strange quark condensate decreases more slowly than the light quark condensate on account of the larger strange quark mass.

As QCD matter expands, cools, and hadronizes, entropy conservation requires that a large number of quarks are produced given that gluons have no quark content and that hadrons are composed of multiple quarks. Simple counting exercises suggest that over two thirds of the constituent quarks in the resulting hadron gas must have been created during hadronization \cite{Bass:2000az}. The energy associated with restoring the quark-antiquark condensates is sufficient to represent much, if not most, of the energy associated with the additional quark production associated with hadronization. Thus, if the condensate were to fluctuate in such a way that 
$\langle \bar{u} u\rangle$ were to differ significantly from $\langle \bar{d} d\rangle$, 
the resulting system should have a strong isospin imbalance.

It is a bit more difficult to motivate iso-vector condensates. In DCC pictures, the iso-singlet scalar condensate is assumed to appear similarly to the iso-vector pseudo-scalar fields (pions) due to chiral symmetry. This symmetry might result in large fluctuations of the fields, or disorientations, and could result in the field being aligned in the wrong directions. For the DIC, one could imagine a similar behavior, but with scalar iso-vector fields.  As we will show below, the cost in energy is significant in the vacuum but decreases dramatically above $T_h$ due to the dissolving quark condensates.  This will facilitate fluctuations, or disorientations of the scalar iso-vector fields with the scalar iso-scalar field, despite the lack of any associated continuous symmetry.

To gain insight into the evolution and possible disorientations of the scalar fields, we consider the $3\times 3$ matrix of scalar fields $M$ whose diagonal elements are 
$\sigma_u = - \langle \bar{u}u \rangle/\sqrt{2}c'$, $\sigma_d = - \langle \bar{d}d \rangle/\sqrt{2}c'$, and $\zeta = - \langle \bar{s}s \rangle/\sqrt{2}c'$.  In terms of conventional fields the relations are $\sigma_u = (\sigma + a_0^0)/\sqrt{2}$ and $\sigma_d = (\sigma - a_0^0)/\sqrt{2}$.
The off-diagonal elements would correspond to charged fields.  From Refs. \cite{JJ,Kapusta:2022ovq} we consider the potential
\begin{eqnarray}
U(M)&=&-\frac{1}{2}\mu^2{\rm Tr}(MM^\dagger)+\lambda({\rm Tr}MM^\dagger MM^\dagger)\\
\nonumber
&+&\lambda'[{\rm Tr}(MM^\dagger)]^2-c({\rm det}~M+{\rm det}~M^\dagger )\\
\nonumber
&-&\frac{c'}{\sqrt{2}}{\rm Tr}[{\cal M}^\dagger M+{\cal M}M^\dagger]\\
\nonumber
&=&-\frac{1}{2}\mu^2(\sigma_u^2+\sigma_d^2+\zeta^2)+\lambda'(\sigma_u^2+\sigma_d^2+\zeta^2)^2\\
\nonumber
&+&\lambda(\sigma_u^4+\sigma_d^4+\zeta^4)
-2c\sigma_u\sigma_d\zeta\\
\nonumber
&-&\sqrt{2}c'(m_u\sigma_u+m_d\sigma_d+m_s\zeta).
\end{eqnarray}
The off-diagonal terms in $M$ have all been set to zero. By setting these to zero, isospin symmetry is broken and fluctuations of the up and down condensates are confined to be along the $I_3=0$ direction.  In what follows we will use numerical values of the parameters as determined in Ref. \cite{Kapusta:2022ovq}

Now we assign a temperature dependence to $\sigma^2(T)=\sigma_u^2(T)+\sigma_d^2(T)$ according to lattice calculations, as shown in Fig. \ref{fig:udsvsT}.  The values of $A$ for the light and strange quarks in physical units are chosen to reproduce the accepted vacuum values of $\sigma_{\rm vac} = f_\pi$ and 
$\zeta_{\rm vac} = (2 f_K - f_\pi)/\sqrt{2}$.  According to the PDG \cite{PDG} the best values are $f_\pi = 92.1$ MeV and $f_K = 110.1$ MeV.  The phenomenology in Ref. \cite{Kapusta:2022ovq} yields $\lambda = 15.01$, $\lambda' = -2.176$, and $\mu^2 = - (472.8\,{\rm MeV})^2$.  The individual values of $c'$ and the light quark $m_q$ and strange quark $m_s$ current quark masses are not renormalization group invariant, only their products are, namely $c' m_q = 8.775\times10^5 \; {\rm MeV}^3$ and $c' m_s = 2.617\times10^7 \; {\rm MeV}^3$.  The parameter $c = 1.732$ GeV.  However, it is expected to have a strong temperature dependence on account of the axial U(1) symmetry being restored at high temperatures.  Therefore, based on instanton calculations \cite{Kapusta:2019ktm}, we assume
$c(T)=c(0)/(1+1.2\pi^2\bar{\rho}^2T^2)^7$ with $\bar{\rho}=0.33$ fm and $c(0) = 1.732$ GeV.

Setting $m_u=m_d \equiv m_q$, the energy minimum occurs when $\sigma_u=\sigma_d$, and the difference of the potential relative to the minimum is
\begin{eqnarray}
\Delta U(T)&=&\frac{1}{2}\lambda \left(\sigma^4-4\sigma_u^2\sigma_d^2 \right)+c(T) \left(\sigma^2-2\sigma_u\sigma_d \right)\zeta\\
\nonumber
&+&\sqrt{2}c'm_q \left(\sqrt{2}\sigma-\sigma_u-\sigma_d \right)\\
\nonumber
&=&\frac{1}{2}\lambda \left[1-\sin^2(2\theta) \right]\sigma^4+c(T) \left[1-\sin(2\theta)\right]\sigma^2\zeta\\
\nonumber
&+&f_\pi m_\pi^2
\left[
1-\frac{\cos\theta+\sin\theta}{\sqrt{2}}
\right]\sigma,
 \end{eqnarray}
where $\sigma_u = \sigma \cos\theta$ and $\sigma_d = \sigma \sin\theta$. For any $\sigma$ the potential minimum is at $\theta=\pi/4$, i.e. $\sigma_u=\sigma_d$. For $\theta=0$, $\sigma_u=\sigma$ and $\sigma_d=0$, and the up-down condensate consists entirely of up quarks, whereas for $\theta=\pi/2$, $\sigma_u=0$ and the condensate consists solely of down quarks.

\begin{figure}
\centerline{\includegraphics[width=0.48\textwidth]{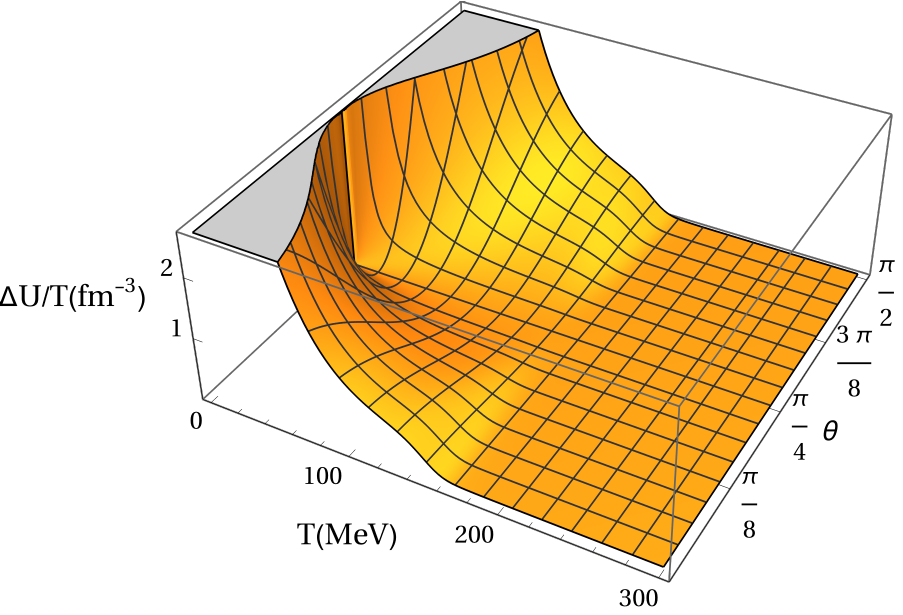}}
\caption{\label{fig:DelU3D}
The energy density penalty divided by the temperature as a function of temperature and angle $\theta$. Large observable effects require fluctuations away from 
$\theta=\pi/4$ at temperatures when quarks are being produced, i.e. $T\approx 160$ MeV.
}
\end{figure}
\begin{figure}
 \centerline{\includegraphics[width=0.48\textwidth]{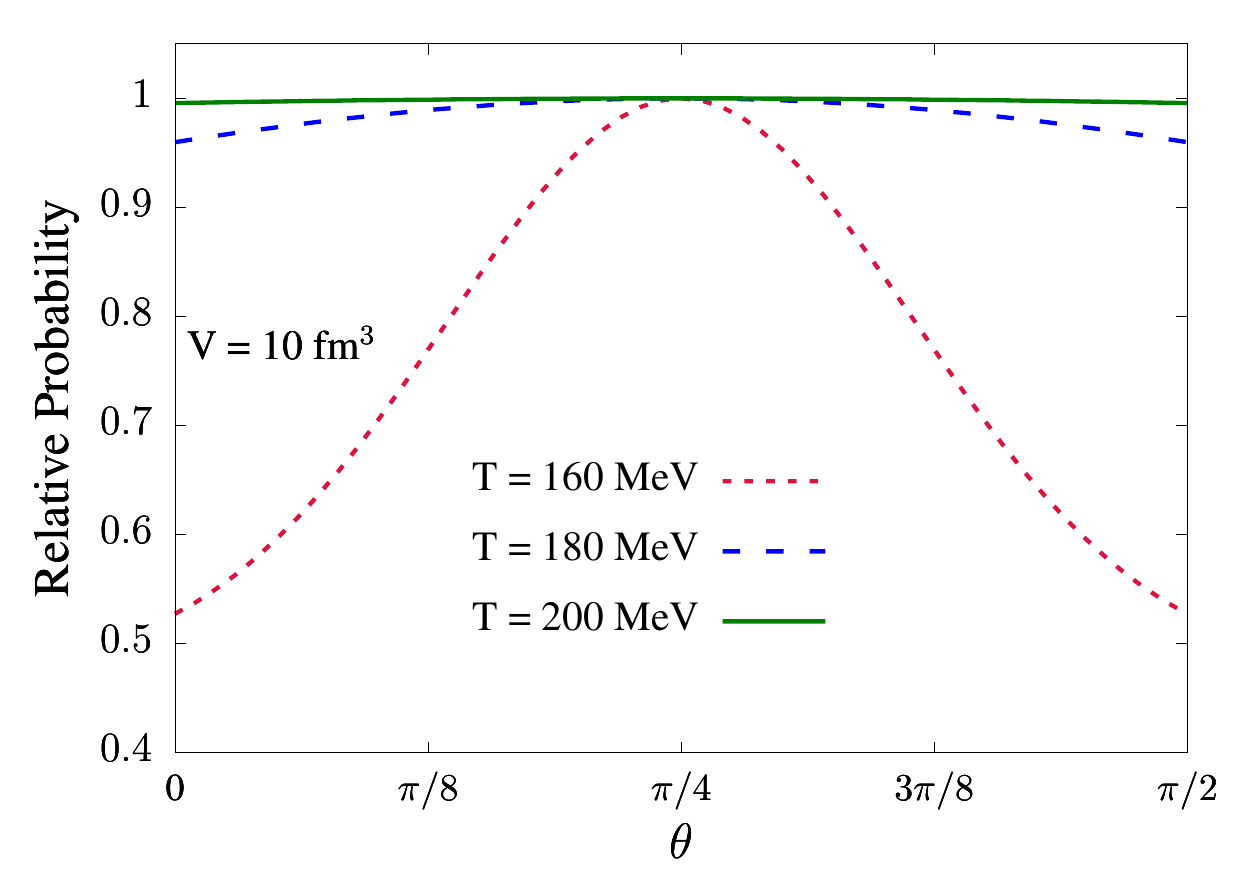}}
\centerline{\includegraphics[width=0.48\textwidth]{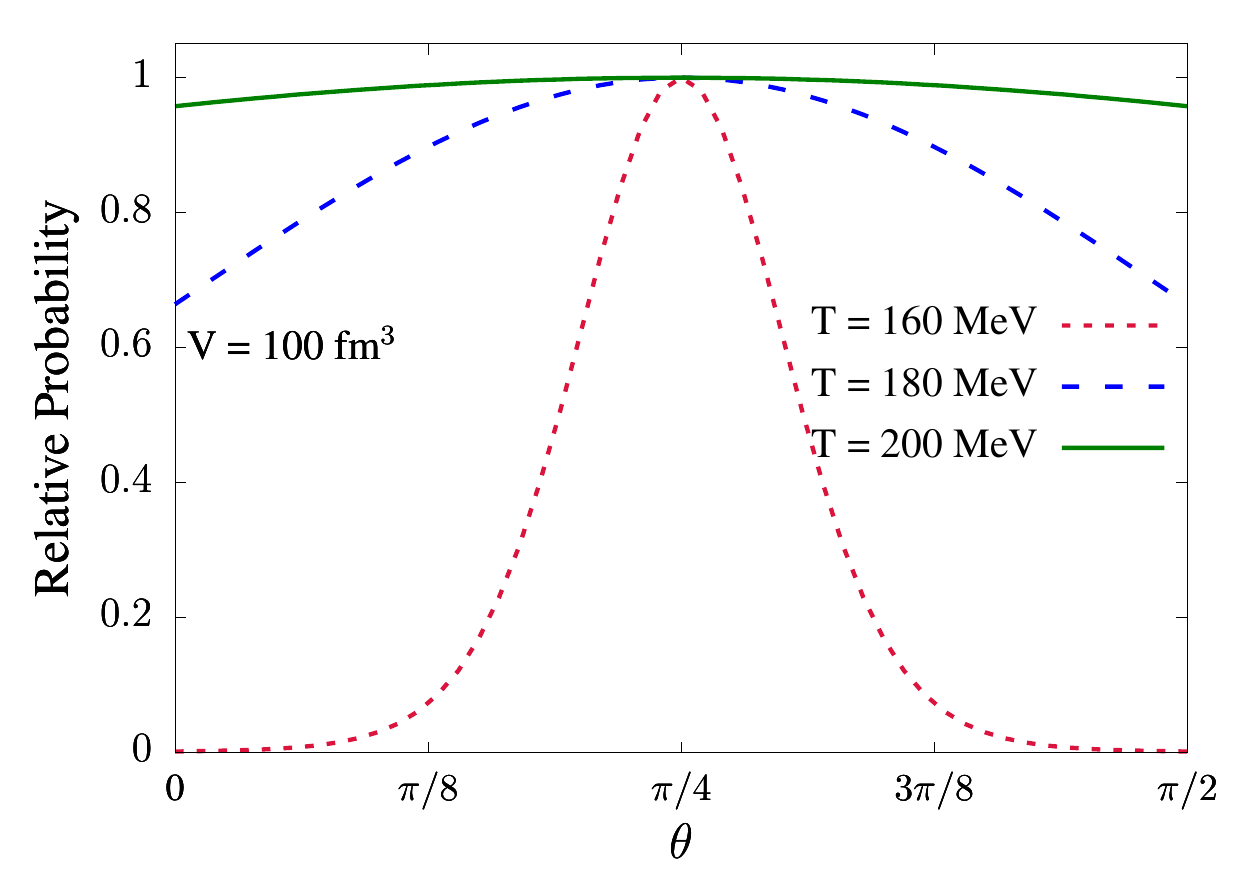}}
\caption{\label{fig:UvsTheta}
The relative probability $P(\theta)={\rm e}^{-V\Delta U/T}$ as a function of $\theta$ for several temperatures. Results are shown for two volumes, $V=10$ fm$^3$ and $V=100$ fm$^3$.
}
\end{figure}

Figure \ref{fig:DelU3D} shows $\Delta U/T$ as a function of temperature and $\theta$. At high temperature there is no penalty because there is no field. As $T$ falls and the fields pick up vacuum expectation values, there is a clear preference for $\theta=\pi/4$. Figure \ref{fig:UvsTheta} presents the relative statistical probability 
${\rm e}^{-V\Delta U/T}$ as a function of $\theta$ for two volumes, $V=10$ fm$^3$ and 100 fm$^3$. From this figure one can see that for temperatures where hadronization takes place, $T\approx 160$ MeV, the pure $\langle \bar{u}u \rangle$ or pure $\langle \bar{d}d \rangle$ condensate is disfavored by a factor of approximately $\thalf$ for a 10 fm$^3$ volume and has vanishingly small probability for the 100 fm$^3$ volume. 

\begin{figure}
\centerline{\includegraphics[width=0.48\textwidth]{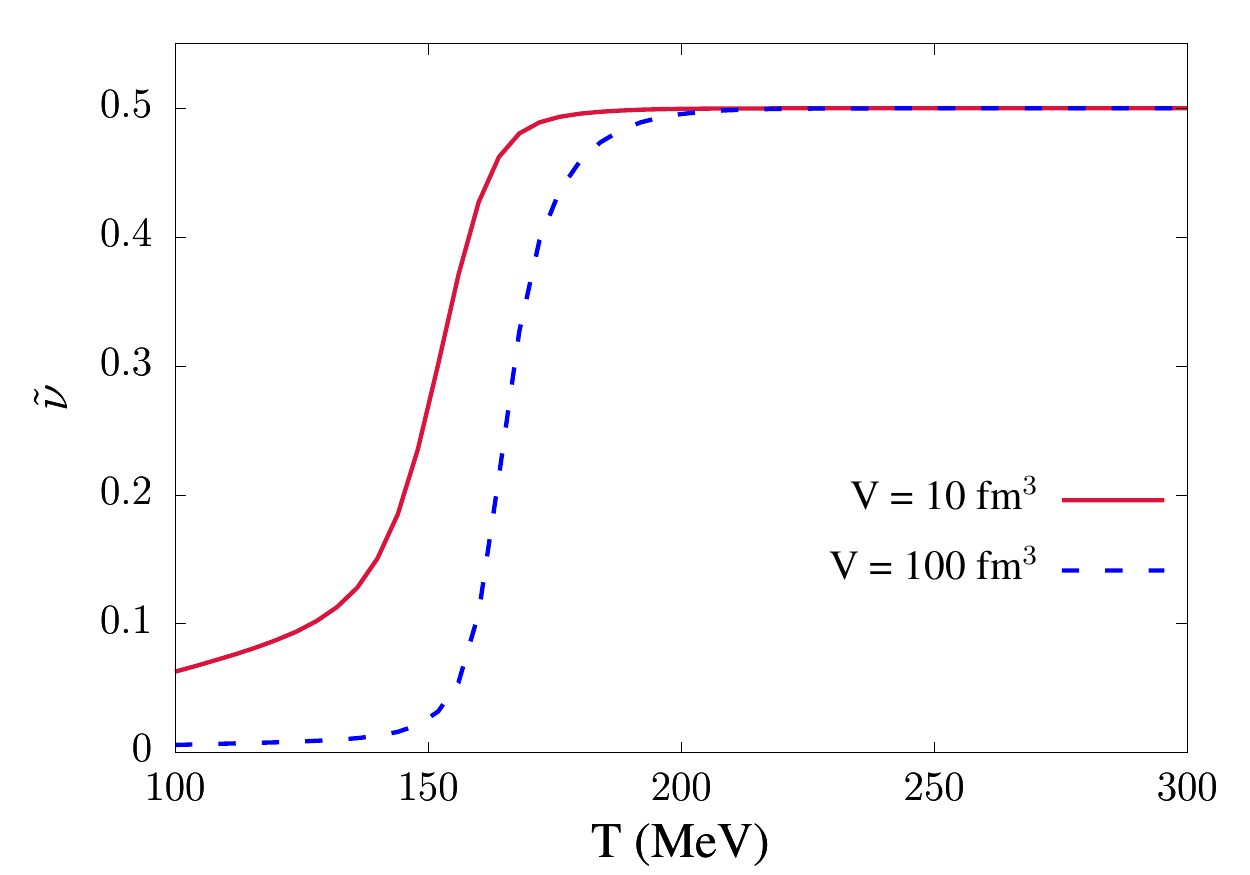}}
\caption{\label{fig:tildenu}
The $\tilde{\nu}$ represents the contribution of any two kaons to $\nu_{\rm dyn}$ if both kaons come from the same domain.  Because larger domains fluctuate less, $\tilde{\nu}$ is significantly smaller for larger volumes.
}
\end{figure}

Fluctuations of $\theta$ lead to fluctuations of the number of up and down quarks, which in turn lead to fluctuations of $\nu_{\rm dyn}$. The phenomenology developed in Ref. \cite{Kapusta:2022ovq} applies here and the data from \cite{ALICE:2021fpb} can be reproduced. Let us assign the fraction of condensate quarks of the up or down flavors as
$f_{u,\bar{u}}=\sigma_u^2/\sigma^2=\cos^2\theta$ and $f_{d,\bar{d}}=\sigma_d^2/\sigma^2=\sin^2\theta$.
For any two kaons from the same domain, i.e. from a region of the same $\theta$, the chance the kaon is charged is $\cos^2\theta$ and the chance the kaon is neutral is $\sin^2\theta$. The probabilities that any two kaons are either both charged or both neutral, minus the probability they have opposite charges is
\be
\tilde{\nu}=\int d\theta~P(\theta)(\cos^4\theta+\sin^4\theta-2\sin^2\theta\cos^2\theta)
\ee
where
\be
P(\theta)=\frac{{\rm e}^{-V\Delta U(\theta)/T}}{\int d\theta'~{\rm e}^{-V\Delta U(\theta')/T}}.
\ee
If $P(\theta)$ is flat then $\tilde{\nu}=\thalf$. For a system with $N_D$ domains, the chance that any two kaons are from the same domain is $1/N_D$.  If the fraction of particles coming from condensates is $F_c$ then
\be
\frac{\nu_{\rm dyn}}{\alpha}=\frac{2}{3}\tilde{\nu}\frac{F_c^2}{N_D}N_{\pm}.
\label{what}
\ee
If one were to split a domain into smaller independent domains, $\tilde{\nu}$ would increase in strength, but the factor of $1/N_D$ would decrease.  Figure \ref{fig:tildenu} shows that stronger signals will emerge from having fewer domains despite the reduction in $\tilde{\nu}$.

Expression (\ref{what}) is similar to that used to estimate the effect of kaon condensates in Ref. \cite{Kapusta:2022ovq}. Let us consider the 10-15\% centrality bin from table III in that reference. If 25\% of the quarks arise from condensates spread over 6 domains, and if there were 125 kaons in the acceptance, one could explain the ALICE measurement if $\tilde{\nu}$ were close to $\oneqt$. Subsequent hadronic interactions might dampen this behavior somewhat. For example, a charged kaon could collide with a pion to form a $K^*$ resonance which might then decay into a neutral kaon. 
The effect would be strengthened if a bigger fraction of quarks came from condensates, or if there were fewer independent domains.  It would also be strengthened if the condensates lagged behind their equilibrium values as the system expanded and cooled due to finite relaxation times.

Given the uncertainties discussed above it is difficult to make precise quantitative predictions.  However, it should be mentioned that the results presented in this letter do not appear to be sensitive to the particular model, as the extended 2+1 model studied in Ref. \cite{Kapusta:2022ovq} yields very similar numbers.  The competing explanations in Refs. \cite{Gavin:2001uk,Kapusta:2022ovq,Nayak:2019qzd} were even more exotic because they relied on 
strange DCC or a finite fraction of kaons occupying a single quantum state, which is difficult to motivate given the low phase space density of kaons, or they relied on fields which are off-diagonal in flavor, such as $\langle \bar{u}s \rangle$.  Unlike the $\langle \bar{u}u \rangle$, $\langle \bar{d}d \rangle$ and $\langle \bar{s}s \rangle$ condensates, such off-diagonal condensates are non-existant in the QCD vacuum. Although the DIC mechanism investigated here is speculative, it seems to be the least questionable explanation for the ALICE measurement of $\nu_{\rm dyn}$ proposed so far.  If one were to measure fluctuations of other pairs of particles, one of which is rich in $u/\Bar{u}$ and the other in $d/\Bar{d}$, such as $\Xi^0$ and $\Xi^-$, it would reinforce the interpretation of the anomaly as due to DIC. This is a prediction of our proposal which would be testable, verifiable, and refutable.

\section*{Acknowledgments}

The work of JK and MS was supported by the U.S. Department of Energy Grant No. DE-FG02-87ER40328, and the work of SP was supported by the U.S. Department of Energy Grant No. DE-FG02-03ER41259.

\end{document}